\documentclass[aps,prl,reprint,amsmath,amssymb,superscriptaddress,longbibliography]{revtex4-2}
\usepackage[utf8]{inputenc}
\usepackage{graphicx}
\usepackage{dcolumn}
\usepackage{bm}
\usepackage{amsmath}

\usepackage{color}
\usepackage{xspace}
\usepackage{xfrac}
\usepackage{color,soul}
\usepackage{graphicx}
\usepackage[colorlinks=true,urlcolor=blue,citecolor=blue,linkcolor=blue,bookmarks=false,pdfstartview={FitH}]{hyperref}
\def\TNS{Ta$_2$NiSe$_5$\,}
\def\TNSS{Ta$_2$NiS$_5$\,}
\def\TNSX{Ta$_2$Ni(Se$_{1-x}$S$_x$)$_5$\,}

\usepackage{color,soul}
\usepackage{enumitem} 

\begin{document}

\title{
 Failed excitonic quantum phase transition in Ta$_2$Ni(Se$_{1-x}$S$_x$)$_5$
}

\author{Pavel A. Volkov}
\email{pv184@physics.rutgers.edu}
\author{Mai Ye}
\email{mye@physics.rutgers.edu}
\affiliation{Department of Physics and Astronomy, Rutgers University, Piscataway, NJ 08854, USA}
\author{Himanshu Lohani}
\affiliation{Department of Physics, Technion - Israel Institute of Technology, Haifa 32000, Israel}
\author{Irena Feldman}
\affiliation{Department of Physics, Technion - Israel Institute of Technology, Haifa 32000, Israel}
\author{Amit Kanigel}
\affiliation{Department of Physics, Technion - Israel Institute of Technology, Haifa 32000, Israel}
\author{Girsh Blumberg}
\email{girsh@physics.rutgers.edu}
\affiliation{Department of Physics and Astronomy, Rutgers University, Piscataway, NJ 08854, USA}
\affiliation{National Institute of Chemical Physics and Biophysics, 12618 Tallinn, Estonia}

\date{\today}

\begin{abstract}
We study the electronic phase diagram of the excitonic insulator candidates Ta$_2$Ni(Se$_{1-x}$S$_x$)$_5$ [x=0, ... ,1] using polarization resolved Raman spectroscopy. 
Critical excitonic fluctuations are observed, that diminish with $x$ and ultimately shift to high energies, characteristic of a quantum phase transition. 
Nonetheless, a symmetry-breaking transition at finite temperatures is detected for all $x$, exposing a cooperating lattice instability that takes over for large $x$. Our study reveals a failed excitonic quantum phase transition, masked by a preemptive structural order.
\end{abstract}
\maketitle

{\it Introduction:} One of the fascinating manifestations of interactions between electrons in solids is the emergence of electronic orders. The fluctuations close to the respective quantum critical points are also believed to be the drivers of a wealth of yet unexplained behaviors, including strange metallicity and high-T$_c$ superconductivity \cite{lohneysen.2007,brando.2016,scalapino2012,shibauchi.2014}.  
In many cases (such as, e.g., nematic \cite{kivelson1998,fradkin2010} or density-wave \cite{grunerbook,kivelson2003} orders) electronic order breaks the symmetries of the crystalline lattice and the corresponding transitions can be, symmetry-wise, identical to structural ones. 
This raises the question of the role of the interplay between electronic and lattice degrees of freedom in the ordering. Even in cases where the lattice only weakly responds to the transition \cite{chu2010,fernandes2014}, the critical temperatures \cite{chu2012,bohmer2014} and quantum critical properties \cite{paul2017} can be strongly modified. Moreover, in a number of cases the origin of the order is still under debate~\cite{comin2016,mou2017,TS2017}, 
as the lattice may develop an instability of its own.

The electronic-lattice dichotomy has recently come to the fore in studies of \TNS \cite{Structure1986,ARPES2009,Transport2017} - one of the few candidate material for the excitonic insulator (EI) phase \cite{neuensch1990,rossnagel2011,du2017,TS2017,li2019Sb,zhu2019graph}. 
EI results from a proliferation of excitons driven by Coulomb attraction between electrons and holes in a semiconductor or a semimetal \cite{Keldysh1965,Ex1967,jerome1967,Ex1968,Ex1970}.  
\TNS exhibits a phase transition at $T_c=328$\,K; while the pronounced changes in band structure \cite{ARPES2009}, transport \cite{Transport2017} and optical \cite{IR2017} properties are consistent with the ones expected for an EI, they allow an alternative interpretation in terms of a purely structural phase transition \cite{subedi2020,ARPES2020,baldini2020,lu2021evolution}. Indeed, EI state in \TNS is expected to break mirror symmetries of the lattice due to the distinct symmetries of the electron and hole states forming the exciton \cite{millis2019}, similar to a structural transition \cite{subedi2020}. Intriguingly, substitution of Se with S has been shown to suppress $T_c$ in transport experiments to zero \cite{Transport2017}, suggesting a possible quantum phase transition (QPT) at $x=x_c$ in \TNSX. Increasing $x$ enhances the band gap in the electronic structure \cite{ARPES2019TNS}, which is known to suppress the EI \cite{jerome1967,Ex1968}, consistent with an EI QPT. On the other hand, the lattice degrees of freedom also evolve with $x$ making it imperative to separately assess the roles of electronic and lattice degrees of freedom throughout the phase diagram of \TNSX.

Experimentally, this is a challenging task. As transitions caused by the lattice and electronic degrees of freedom break the same symmetries  \cite{millis2019,subedi2020}, their signatures may appear identical in thermodynamic (e.g., specific heat \cite{Transport2017}) and symmetry-sensitive (X-ray diffraction \cite{Structure1986}) probes, as well as in the single-electron spectra \cite{ARPES2009,ARPES2020}. Probing the collective dynamics out of equilibrium could provide more information \cite{Fast2018a,baldini2020,lu2021evolution}, but raises the question of whether the non-equilibrium state preserves the interaction between the structural and electronic modes intact \cite{Ye2021}. Finally, due to the even-parity nature of the critical mode \cite{volkov2020}, the dipole selection rules forbid direct observation of the order parameter response in optical absorption \cite{IR2017}. A promising technique to address the near-equilibrium collective dynamics is polarization-resolved Raman scattering \cite{volkov2020,kim2021,blumberg2021comment}, that also allows to detect symmetry breaking  independently \cite{kung2015,ye2019yb,shangfei2020}. Furthermore, analysis of the Raman data  \cite{volkov2020,Ye2021} enables to deduce the individual contributions of the lattice and electronic modes to the transition, making this technique unique in its scope.
	
In this Letter, we use polarization-resolved Raman scattering to study the dynamics of electronic excitations throughout the phase diagram of \TNSX. We reveal the presence of low-energy excitonic modes that soften on cooling towards $T_c(x)$. 
This softening indicates, that in the absence of lattice effects, a purely excitonic transition would have taken place at $T_{ex}(x)$, which we deduce to be smaller than $T_c(x)$. 
On increasing sulfur content $x$, $T_{ex}(x)$ is suppressed to negative values and for $x=1$ low-energy excitons are no longer observed, as expected for an excitonic insulator quantum phase transition. 
However, the actual $T_c(x)$ remains finite for all $x$, implying the presence of a cooperating lattice instability, obscuring the suppression of the excitonic order. The study thus reveals a "failed" excitonic quantum phase transition in \TNSX masked by a preemptive structural order, that takes over as the electronic instability is suppressed.

\begin{figure*}
	\centering
	\includegraphics[width=\linewidth]{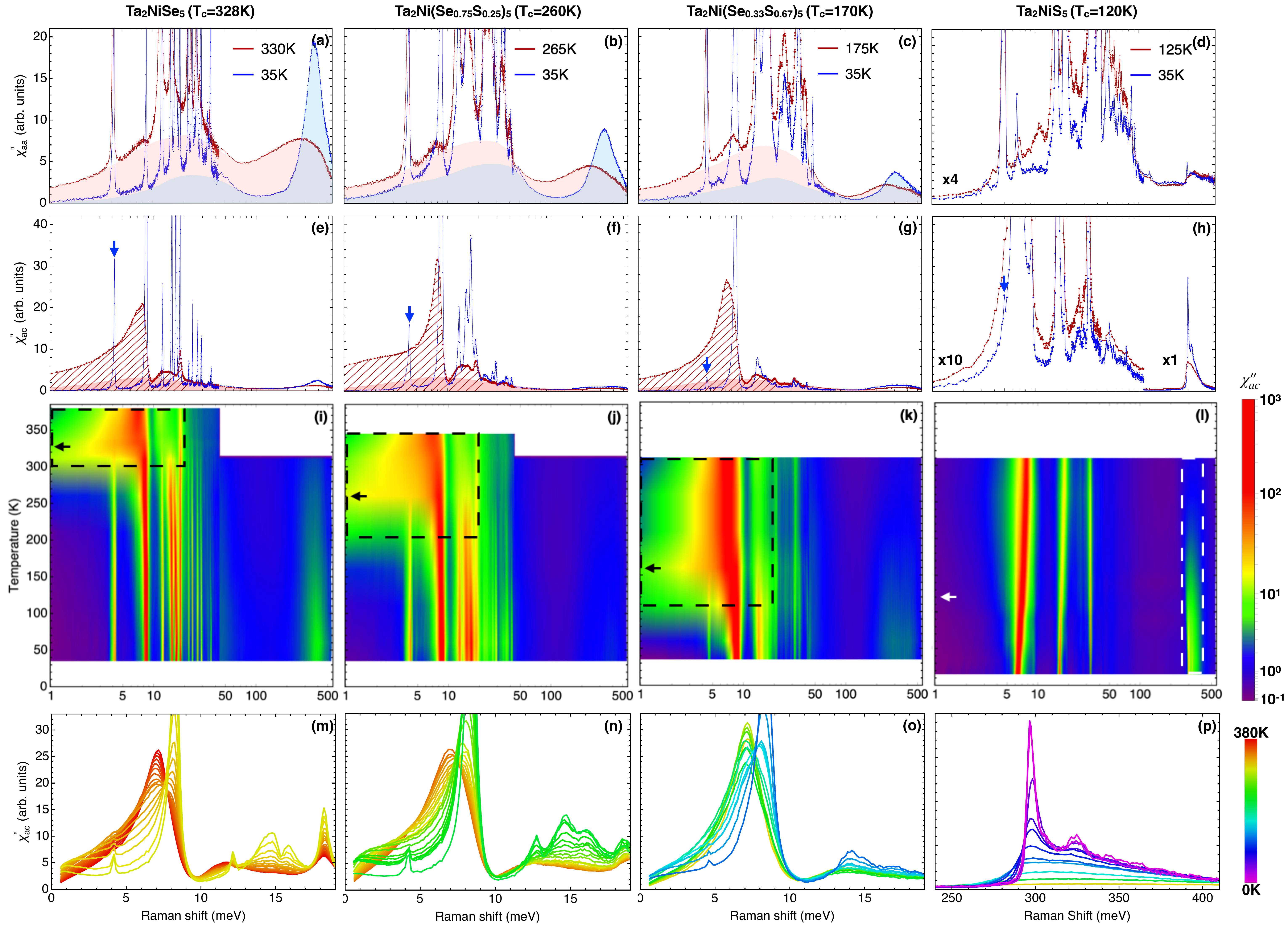}
	\vspace{-4mm}
	\caption{Overview of the polarization-resolved Raman response $\chi''(\omega,T)$ in Ta$_2$Ni(Se$_{1-x}$S$_x$)$_5$. (a-d) Response in $aa$ polarization geometry corresponding to fully symmetric ($A_g$) excitations above (red) and below (blue) $T_c$ for $x=0-1$. 
	Shading highlights the bare electronic contribution to the response. 
	(e-f) Same for $ac$ geometry, probing the excitations with the symmetry of the order parameter ($B_{2g}$) for $T>T_c$ (red). Unlike $aa$ geometry, phonons show an extremely anisotropic Fano lineshape (hatching), indicating their strong interaction with the electronic continuum (red shading). For $T<T_c$ (blue) excitations observed in $aa$ geometry above $T_c$ appear (arrow) due to symmetry breaking. (i-l) Temperature dependence of $\chi''_{ac}(\omega,T)$; an enhancement at low energies near $T_c$ (arrow) is observed for $x=0,0.25,0.67$. For $x=1$ no low-energy response is present. (m-p) Details of  $\chi''_{ac}(\omega,T)$ for regions marked by dashed lines in (i-k). (m-o) Fano lineshape of low-energy phonons due to interaction with electronic continuum. (p) High-energy peak due to an uncondensed exciton in \TNSS.
	}
	\vspace{-4mm}
	\label{fig:overview}
\end{figure*}

{\it Experimental:} We performed Raman scattering experiments on single crystals \TNSX with varying Se/S content, grown using the chemical vapor transport (CVT) method~\cite{Ye2021}\footnote{See Supplementary Material at [URL will be inserted by publisher] for the details of crystal growth and characterization, Raman measurements and analysis, which includes Refs. \cite{Transport2017,nakano2018,Ye2021,Klemens1966,Raman2019,Hayes2004,kung2015,fano1961,klein1983,goldenfeld2018,Schuller2006,yusupov2010,zong2019,volkov2020,elliot1957,haug2009quantum,toyozawa1964,segall1968}.}. 
The measurements were performed in a quasi-back-scattering geometry on samples cleaved to expose the $ac$ crystallographic plane with the 647\,nm line from a Kr$^+$ ion laser excitation, details presented in Ref.\,\cite{Ye2021}. 
The selection rules in the high-temperature orthorhombic (point group $D_{2h}$) phase imply that $ac$ polarization geometry probes excitations with $B_{2g}$ symmetry (same as that of the order parameter), while $aa$ geometry probes the fully symmetric $A_g$ ones~\cite{volkov2020,kim2021,Ye2021,blumberg2021comment}. 
Below $T_c$, the point group symmetry is reduced to $C_{2h}$ and the two irreducible representations merge, such that excitations from $ac$ geometry above $T_c$ may appear in $aa$ geometry and vice versa. Their appearance allows to determine $T_c$ from the Raman spectra.
 
{\it Data overview:} Summarized temperature dependence of the Raman susceptibility $\chi''(\omega,T)$ is presented in Fig.\,\ref{fig:overview}. The samples with $x=0,0.25,0.67$ show qualitatively similar spectra. At low energies, phonon peaks are observed on top of a smooth background, which we attribute to electronic excitations. On cooling, a pronounced redistribution of electronic intensity in a wide range of energies is observed, leading to a formation of a gap-like suppression followed by a high-energy feature Fig.\,\ref{fig:overview}(a)-(c).
This feature at 380\,meV for \TNS has been attributed to the coherence factors at the gap edge of an EI~\cite{volkov2020}. In $ac$ geometry, a pronounced enhancement at low energies is evident close to $T_c$, consistent with critical mode softening near a second-order phase transition, Fig.\,\ref{fig:overview}(i)-(k). In the same temperature region, the lineshapes of the low-energy phonons show strongly asymmetric Fano form (Fig.\,\ref{fig:overview}(m)-(o)) - a known signature of interaction with an electronic excitation continuum \cite{volkov2020,Ye2021,klein1983}. This indicates the presence of low-energy symmetry-breaking electronic excitations, that soften close to $T_c$. At low temperatures, the asymmetry disappears (Fig.\,\ref{fig:overview}(e)-(g)), a behavior consistent with a gap opening in an EI. 
On increasing sulfur content $x$, the temperature where the strongest low-energy enhancement is observed progressively lowers (Fig.\,\ref{fig:overview}(i)-(k)), 
and the $A_g$-symmetry feature at about 380\,meV moves slightly to lower energy and becomes less pronounced.

The signatures for Ta$_2$NiS$_5$~($x=1$) are rather different: in the $ac$ geometry, low-energy electronic excitations are absent at all temperatures, indicating the presence of a direct gap. 
This implies that between $x=0.67$ and $x=1$ the electronic structure undergoes a Lifschitz transition from a semimetallic to an insulating one. 
The intensity in the $aa$ geometry at low energies is also pronouncedly smaller than for the other samples and no broad high-energy peak is observed at low temperatures. 
On the contrary, a sharp feature at about $0.3$\,eV emerges in $ac$ geometry on cooling below 100\,K.

\begin{figure}
	\centering
	\includegraphics[width=0.8\linewidth]{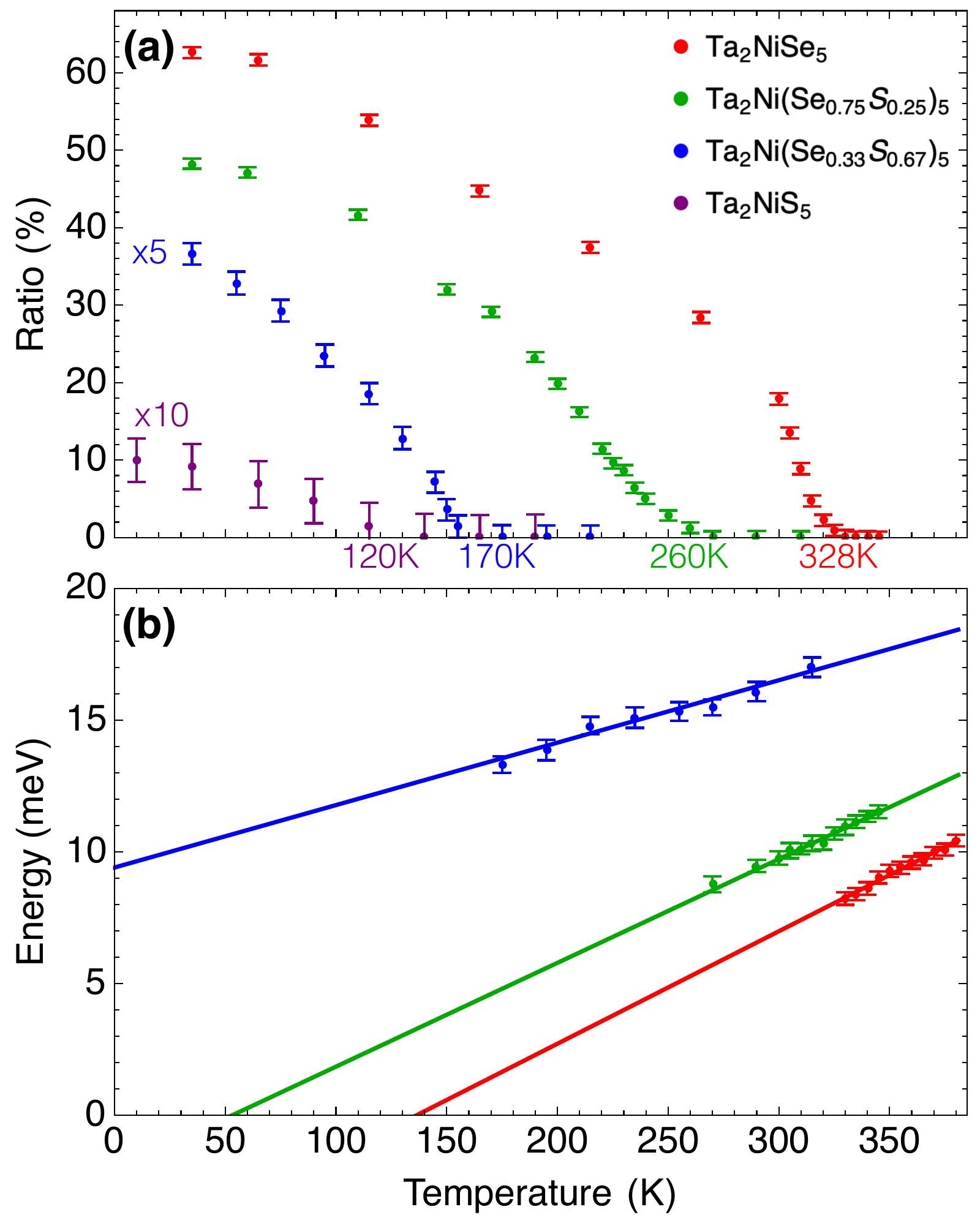}
	\vspace{-4mm}
	\caption{
	The parameters deduced from the Raman response data, Fig.\,\ref{fig:overview}. 
	(a) Integrated 'leakage' intensity into $ac$ scattering geometries of the lowest-energy A$_g$ phonon mode labeled in panels Fig.\,\ref{fig:overview}(e-h), normalized by the intensity in the dominant $aa$ geometry. 
	The mode's appearance in the $ac$ scattering geometry below $T_c$ implies the onset of symmetry breaking.
	(b) Excitonic energy $\Omega_e(T)$, Eq.\,\eqref{eq:dyn} obtained from the Fano fits to the lineshapes in Fig.\,\ref{fig:overview}.
	Lines represent linear fits to the points.}
	\label{fig:param}
	\vspace{-6mm}
\end{figure}

\begin{figure}
	\centering
	\includegraphics[width=0.85\linewidth]{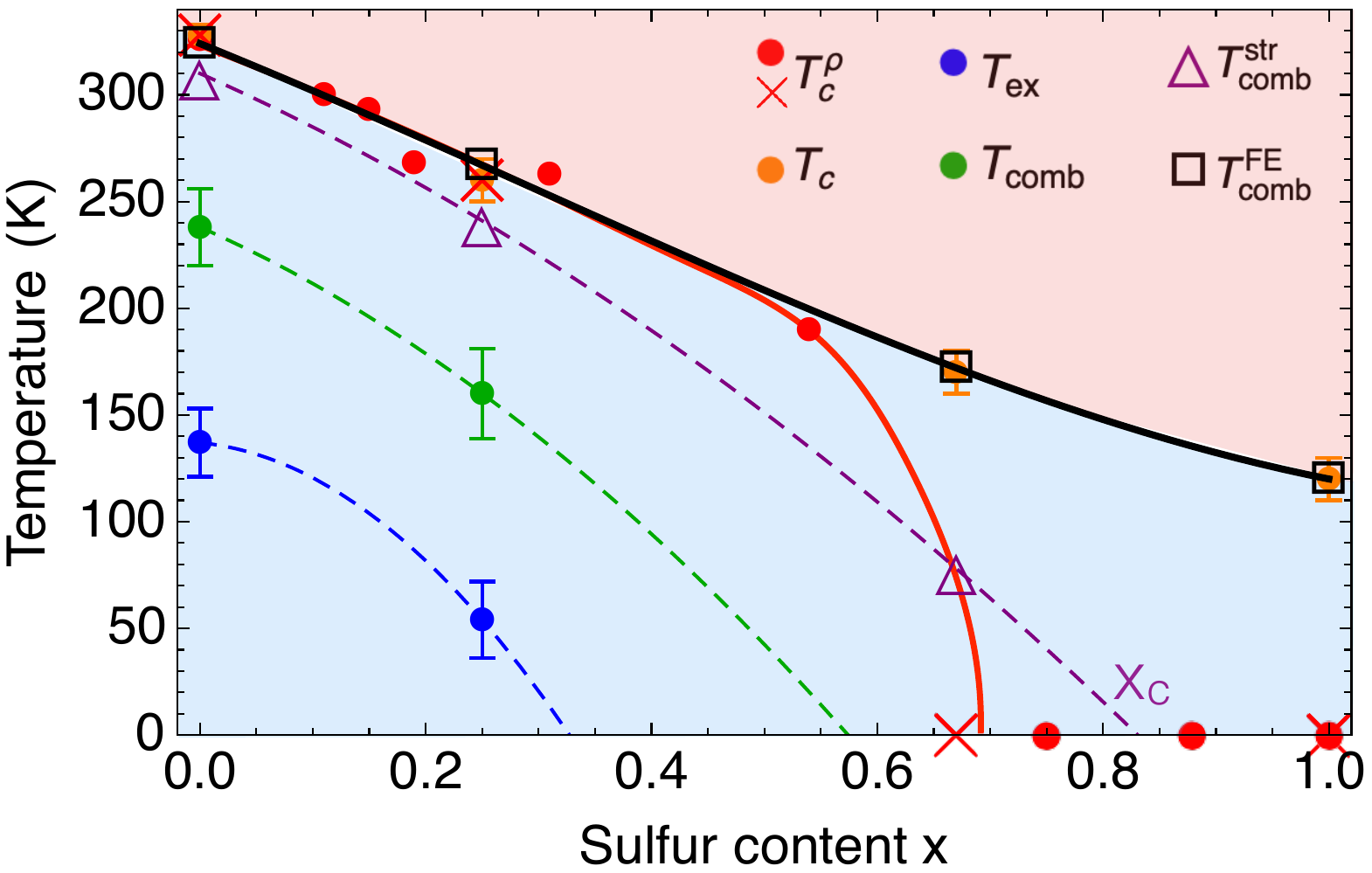}
	\vspace{-4mm}
	\caption{
		Phase diagram of Ta$_2$Ni(Se$_{1-x}$S$_x$)$_5$. Orange points: the 
		symmetry breaking transition temperature $T_c(x)$ obtained from the onset of phonon intensity 'leakage', Fig.\,\ref{fig:param}(a).
		Red points and crosses: $T_c^{\rho}(x)$ adapted from transport studies Refs.\,\cite{Transport2017,Ye2021}.
		For low sulfur concentration $x$, the soft 	excitonic mode, Fig.\,\ref{fig:param}(b), would drive the transition at temperature $T_{ex}(x)$ (blue points), that is enhanced to $T_{comb}(x)$ by coupling to inert optical phonons (green points), and is further enhanced to $T_{comb}^{str}(x)$ by coupling to the $B_{2g}$ strain (purple triangles). 
		For large $x$, the excitonic softening is suppressed, while a ferroelastic instability leads to a finite $T_{comb}^{FE}(x)$ (black squares). 
		In the absence of the lattice instability, a lattice-shifted electronic QPT would have occurred at $x_c$ (dashed purple line). Additionally, in the same proximity, the band structure undergoes semimetal-to-semiconductor Lifschitz transition (see text).
	}
	\vspace{-5mm}
	\label{fig:phdiag}
\end{figure}

{\it Symmetry-breaking transition:} We address first the presence of a phase transition by studying the appearance of new modes in the broken-symmetry phase, as outlined above.
In Fig.\,\ref{fig:param}(a) we show the temperature dependence of such a 'leakage' phonon intensity marked by arrow in Fig.\,\ref{fig:overview}(e)-(h). One can see the appearance of the 'leaked' intensity below $T_c$ in the pure Se case, as well as the decrease of $T_c$ with S doping. 'Leakages' of other modes appear below the same temperature $T_c$~\cite{Ye2021,Note1}. 
At low $x$ the obtained values of $T_c$ agree with the ones deduced from transport and specific heat measurements~\cite{Transport2017,Ye2021}, Fig.\,\ref{fig:phdiag}. 
However, in contrast to the transport data reported in  Ref.\,\onlinecite{Transport2017}, we find that the symmetry-breaking transition persists for all compositions, although the 'leakage' intensity is strongly suppressed with higher $x$. 
The latter suggests that the phase transition signatures in thermodynamic and transport measurements may become too weak to be observed at large $x$, especially since the system becomes more insulating with $x$. 
The structural signatures, e.g. the deviation of the monoclinic angle $\beta$ from 90$^\circ$, should be also strongly suppressed, being already weak at $x=0$ \cite{Structure1986}.

The phase transition for $x=1$, where no low-energy softening is observed (Fig.\,\ref{fig:overview}(l)), indicates a different transition mechanism. Below we analyze our data to elucidate the origin of the transition as a function of $x$.

{\it Electronic contribution to the phase transition:} 
We investigate first the soft-mode behavior observed for $x \leq 0.67$ (Fig.\,\ref{fig:overview}(i-k)).
In particular, we analyze the asymmetric lineshapes of the low-energy part of $\chi_{ac}''(\omega,T)$ around $T_c$ (Fig.\,\ref{fig:overview}(m-o)) using an extended Fano model~\cite{volkov2020,Ye2021}. The model assumes three phononic oscillators (which is the number of $B_{2g}$ modes in the orthorhombic phase) interacting with a continuum of excitonic origin. The latter is expected to arise from the excitonic fluctuations in a semimetal, overdamped due to the allowed decay into particle-hole pairs. 
Close to the transition, the dynamics of the excitonic mode is governed by the time-dependent Landau equations~\cite{goldenfeld2018,Schuller2006,yusupov2010,zong2019}. 
Together with the standard oscillator dynamics of the phonons the system is described by
\begin{equation}
\begin{gathered}
\{\partial_t+\Omega_e(T) \} \varphi +\sum_{i=1}^3 \tilde{v}_i \eta_i =0,
\\
\{\partial^2_t+2\gamma_i(T) \partial_t+ \omega_{pi}^2(T) \}\eta_i + \tilde{v}_i \varphi =0,
\end{gathered}
\label{eq:dyn}
\end{equation}
where $\eta_{i=1,2,3}$ and $\varphi$ are the collective coordinates (order parameters) of the optical phonons and excitons, respectively. $\Omega_e(T)$ is the characteristic energy of the excitonic fluctuations, $\omega_{pi}(T)$ and $\gamma_i(T)$ are the phonon frequencies and scattering rates, and a bilinear exciton phonon-coupling $\tilde{v}_i$ is assumed. The linear response of the system Eq.\,\eqref{eq:dyn} determines the Raman susceptibility. 
The resulting model~\cite{Note1} is a generalization of the standard Fano model~\cite{klein1983} for Raman scattering in metals to the case of three phonons and the continuum response determined from the Landau theory, Eq.\,\eqref{eq:dyn}. 
The purely excitonic part of the response has then the form of a broad continuum  $\chi''_{cont}(\omega,T) \propto \frac{\omega}{\Omega_e^2(T)+ \omega^2}$, in contrast to the Lorentzian phonon peaks. 
The interaction between the phonons and the excitonic continuum leads to an asymmetric broadening of the peaks \cite{Ye2021}, allowing to capture the observed lineshapes in great detail \cite{Note1,volkov2020,Ye2021}.

We now discuss the parameters deduced from the Fano model fits. 
The phonon frequencies $\omega_{pi}(T)$ do not soften near $T_c$  \cite{Ye2021}, ruling out a zone-center phonon instability \cite{subedi2020}. 
On the other hand, $\Omega_e(T)$ (Fig.\,\ref{fig:param}(b), solid lines) consistently softens above $T_c$ for all semimetallic samples. The linear temperature dependence $\Omega_e(T)\sim T-T_{ex}$ implies that a purely electronic transition would have taken place at $T_{ex}<T_c$ for $x=0,0.25$ (Fig. \ref{fig:phdiag}, blue symbols). 
The strongly negative $T_{ex}$ for $x=0.67$ indicates that the exciton softeninng alone would not have lead to a transition at this sulfur concentration.

The suppression of the excitonic instability with $x$ is even more evident in \TNSS, Fig.\,\ref{fig:overview}(d,h,l), where the low-energy electronic response is altogether absent  due to a direct band gap \cite{ARPES2019TNS}. 
Instead, we observe a sharp $B_{2g}$-symmetry mode at $0.3$\,eV, Fig.\,\ref{fig:overview}(p), consistent with an in-gap exciton. 
It is followed by a weaker feature at $0.325$\,eV and an intensity 'tail' at higher energies up to around $0.4$\,eV. 
The natural interpretation of the second peak is the second state of the Rydberg series (i.e. 2S exciton), while the high-energy intensity 'tail' can be attributed, in analogy with optical absorption spectroscopy, to the Rydberg states of higher order and interband transitions~\cite{elliot1957,haug2009quantum} with possible contributions from phonon-assisted exciton transitions~\cite{toyozawa1964,segall1968}. 
A 'leakage' of the exciton features is also observed in $aa$ geometry due to symmetry breaking, Fig.\,\ref{fig:overview}(d)~\cite{Note1}. 
On heating, all the features broaden and eventually smear out above $100$\,K. The increase of the linewidth of the excitonic features can be attributed to the interaction with acoustic and optical phonons~\cite{rudin1990}.

{\it Ferroelastic transition in \TNS:} 
The presented observations show that the excitonic instability on its own cannot explain the occurrence of a transition for samples with large $x$, calling for a more careful consideration of the lattice effects. The most vivid is the case of \TNSS, where the excitonic response is confined to high energies (Fig.\,\ref{fig:overview}(p)). Three $B_{2g}$ optical phonon modes (Fig.\,\ref{fig:overview}(l)), on the other hand, exhibit some (around 15\% maximum) softening on cooling. However, their energies never soften below 6.5\,meV, nor exhibit anomaly at the transition temperature $120$ K. We note that the number of $B_{2g}$ modes is restricted to three by the space group of the orthorhombic \TNSS, implying the absence of any other $B_{2g}$ optical modes beyond those shown in Fig.\,\ref{fig:overview}(l). Consequently, an instability of zone center phonons in \TNSS \cite{windgatter2021common} is ruled out by the data.

The only remaining option for the transition origin in \TNSS is an instability of the acoustic modes, i.e. ferroelasticity~\cite{ferroelast2012}\footnote{Not to be confused with ferroelectricity, usually driven by a soft optical mode \cite{cochran.1960}.}, driven by softening of the $B_{2g}$ shear modulus $C_{ac}(T)$. Indeed, the acoustic modes are not observed directly in Raman due to their extremely low energies and weak coupling to light~\cite{Ye2021,gallais2016}. However, we will show now that the effects of the ferroelastic instability can be observed at $x<1$ via its coupling to the low-energy excitons.

{\it Electronic-structural phase diagram:} We will now demonstrate that the entire phase diagram of \TNSX can be understood by including the interaction between excitonic and lattice modes. As has been noted above, the bare excitonic transition temperature $T_{ex}$ (Fig.\,\ref{fig:phdiag}, blue line) is significantly lower than the actual $T_c$. However, even the coupling of excitons with the otherwise inert optical phonons can affect the transition temperature. For a coupled excitonic-optical phonon system, the transition temperature $T_{comb}(x)$ corresponds to the appearance of a zero-energy solution of Eq.\,\eqref{eq:dyn} deduced from \cite{Note1}:
\begin{equation}
\Omega_e(x,T_{comb}(x)) -\sum_i \frac{\tilde{v}_i^2}{\omega_{pi}^2(x,T_{comb}(x))}=0, 
\end{equation}
where all the parameters of this equation are deduced from the Fano analysis of the Raman data, Fig.\,\ref{fig:overview}(m-o), following Ref. \onlinecite{Ye2021}. The resulting temperature $T_{comb}(x)$ is shown in Fig.\,\ref{fig:phdiag} (green line). While higher than $T_{ex}(x)$, there is still discrepancy with $T_c$: for example, $T_{comb}(x=0.67)$ is negative, while the actual $T_c(x=0.67)$ is 170\,K.

We now include the effects of coupling of the excitonic order parameter $\varphi$ to the $B_{2g}$ strain $\varepsilon_{ac}$ (acoustic modes). A linear coupling between the two is allowed by symmetry \cite{chu2012,bohmer2014,volkov2020} and leads to a further modified equation for the transition temperature \cite{volkov2020,Note1}:
\begin{equation}
\begin{gathered}
\Omega_e(x,T_{comb}^{FE}(x)) -\sum_i \frac{\tilde{v}_i^2}{\omega_{pi}^2(x,T_{comb}^{FE}(x))}
\\
- \lambda^2/[2C_{ac}(T_{comb}^{FE})]=0,
\end{gathered}
\label{eq:tcstr}
\end{equation}
where $\lambda$ and $C_{ac}(T)$ is the strain-exciton coupling constant and the $B_{2g}$ shear modulus, respectively. 
To capture the ferroelastic instability at $x=1$, we assume a Curie-Weiss behavior of the shear modulus $C_{ac}^{-1}(T) = C_{ac(0)}^{-1}+\frac{a}{T-120 \text{K}}$. Using $\lambda^2 C_{ac(0)}^{-1}$ and $\lambda^2 a$ as fitting parameters, the observed $T_c(x)$ can be captured very accurately, see the black line in Fig.\,\ref{fig:phdiag}. Importantly, the effects of the ferroelastic softening become noticeable well before $x=1$. The purple dashed line $T_{comb}^{str}(x)$ in Fig.\,\ref{fig:phdiag} shows the transition temperature $T_{comb}^{str}(x)$ obtained ignoring ferroelastic softening (i.e., taking $C_{ac}^{-1}(T) = C_{ac(0)}^{-1}$ in Eq. \eqref{eq:tcstr}). The result deviates strongly from actual $T_c(x)$ already at $x=0.67$. Continuing the trend further suggests a complete suppression of ordering at $x_c\approx0.8$ in the absence of ferroelasticity. At the same time, for low $x$, $T_{comb}^{str}(x)$ and $T_{comb}^{FE}(x)$ are almost indistinguishable, suggesting that ferroelasticity does not play a role in that case.

This picture bears important consequences for the physics of \TNSX. At low $x$, the transition is driven, to a good approximation, only by the excitonic softening.  
On increasing $x$, the lattice softening becomes more important, and for $x=1$ the transition is purely ferroelastic. In the absence of ferroelasticity, an electronic QPT would have occurred at $x_c\approx 0.8$. We remind that the excitonic and lattice orders break the same symmetries in \TNSX. Consequently, no change in symmetry occurs as a function of $x$ at low temperatures and a true QPT at $T=0$ is avoided (unlike the case of superconductivity emerging near quantum critical points). However, this does not preclude critical electronic fluctuations, associated with the ``failed" QPT at $x_c\approx 0.8$ to be observed at sufficiently high temperatures (higher than the bare lattice transition temperature of 120\,K) \cite{paul2017}.

The presence of quantum critical fluctuations due to a "failed" excitonic QPT lends a natural explanation to the signatures of strong correlations observed in \TNS. In particular, a filling-in, rather then closing of the gap in $aa$ Raman spectra has recently been connected to strong electronic correlations \cite{volkov2020}; moreover, ARPES studies \cite{ARPES2014} suggest the presence of "preformed excitons" well above $T_c$ also characteristic of correlated regime. Similar temperature evolution of $aa$ spectra is also observed for the doped samples, Fig.\,\ref{fig:overview}(b,c). 
Interestingly, while the intensity of the coherent $aa$ peak is suppressed with doping, as is expected from mean-field theory \footnote{In a two-band model of EI $\chi''_{aa}(\omega)\sim \nu_0\sqrt{2W/(\omega-2W)}$ \cite{volkov2020}, where $\nu$ is the density of states above $T_c$ and $W$ is the interband hybridization, which is the order parameter for pure EI \cite{millis2019}.}, the position of the peak changes only weakly. 
The latter behavior indicates strong correlations which get a natural explanation in terms of the quantum critical fluctuations from the "failed" QPT.
Finally, ferroelasticity may be suppressed by strain~\cite{ikeda2018} or pressure~\cite{Transport2017} raising the possibility to reveal the bare EI QPT at low temperatures. For a semimetallic band structure, the EI QPT has been predicted to lead to non-Fermi liquid behavior \cite{exQCP1}, mass enhancement \cite{exQCP2} or emergence of superconductivity~\cite{volkov2018}. Interestingly, superconducting dome near the end point of the monoclinic phase has been recently reported in \TNS under pressure \cite{matsubayashi2021}.

{\it Conclusions:} In this work, we used polarized Raman scattering to study the phase diagram of the excitonic insulator candidates \TNSX, disentangling the roles of structural and electronic ordering. 
We revealed a "failed" excitonic insulator quantum phase transition at $x_c\approx0.8$, avoided due to a lattice instability below 120\,K. At low sulfur content $x$ we observed a soft excitonic mode driving the transition, while at large $x$ this mode ultimately transforms into a high-energy exciton, unable to drive the transition. We further exclude the instability of optical phonons and demonstrate that a 
ferroelastic instability yields an explanation of the observed symmetry-breaking transition. While the excitonic quantum phase transition is avoided due to the low-temperature lattice instability, the associated critical fluctuations can still be present at high temperatures \cite{paul2017}, explaining the correlation effects in \TNS. Furthermore,	selective control of the structural and excitonic instability by strain~\cite{ikeda2018} or pressure~\cite{Transport2017} can turn \TNSX into a platform to study the excitonic QPT at low temperatures as well as quantum critical ferroeleasticity~\cite{paul2015}.

\begin{acknowledgments}
Authors acknowledge discussions with K.\,Haule.  
The spectroscopic work conducted at Rutgers (M.Y. and G.B) was supported by the National Science
Foundation (NSF) Grants No. DMR-1709161 and No. DMR-2105001. P.A.V. acknowledges the Postdoctoral Fellowship support from the Rutgers University Center for Materials Theory. 
The sample growth and characterization work conducted at the Technion was supported by the USA-Israel Binational
Science Foundation (BSF) Grant No. 2020745 and by the Israel Science Foundation Grant No. 1263/21 (H.L., I.F. and A.K.). 
H.L. was supported in part by a PBC fellowship of the Israel Council for Higher Education. 
The work at NICPB was supported by the European Research Council (ERC) under Grant Agreement No. 885413. 

P.A.V and M.Y. contributed equally to this project. 
\end{acknowledgments}


%

\newpage

\clearpage
\onecolumngrid
\appendix
\renewcommand{\thefigure}{S\arabic{figure}}
\addtocounter{equation}{-2}
\addtocounter{figure}{-3}
\renewcommand{\theequation}{S\arabic{equation}}
\renewcommand{\thetable}{S\arabic{table}}

\begin{center}
	\textbf{\Large
		Supplemental Material for: \\ Failed excitonic quantum phase transition in Ta$_2$Ni(Se$_{1-x}$S$_x$)$_5$}
\end{center}

\section{Experimental}

\subsection{Sample preparation and characterization}

Single crystals of Ta$_2$Ni(Se$_{1-x}$S$_x$)$_5$ family were grown by chemical vapor transport method. Elemental powders of tantalum, nickel, selenium and sulfur were mixed with stoichiometric ratio and then sealed in an evacuated quartz ampule with a small amount of iodine as the transport agent. The mixture was placed in the hot end of the ampule ($\sim$950$^\circ$C) under a temperature gradient of about 10$^\circ$C/cm. After about a week mm-sized needle-like single crystals were found at the cold end of the ampule. The crystals are shiny and cleave easily. We used x-ray diffraction (XRD) and electron dispersive X-ray spectroscopy to verify the exact composition of the crystals and their uniformity.

\begin{figure}[b]
	\centering
	\includegraphics[width=0.4\linewidth]{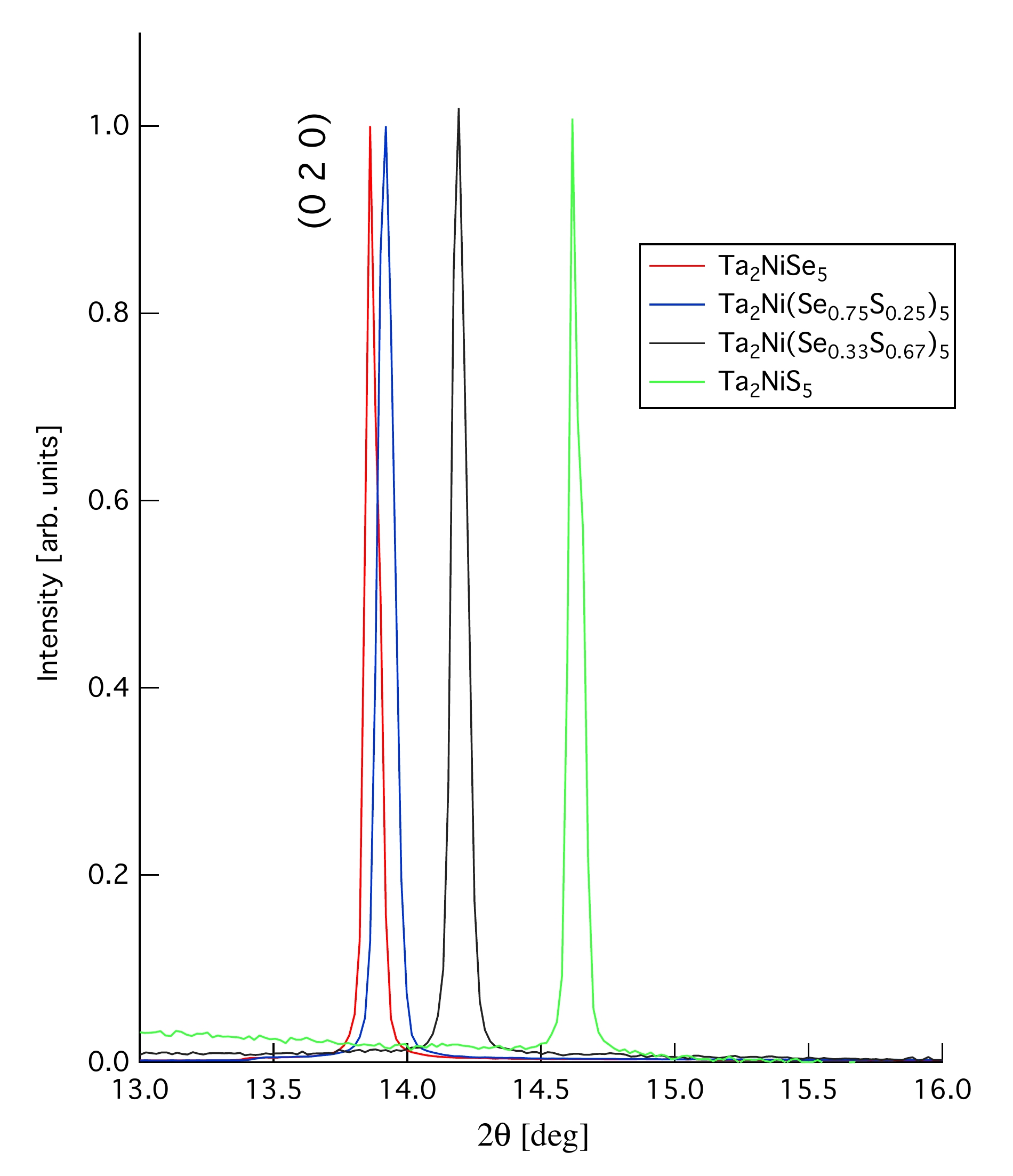}
	\caption{
		Typical X-ray $2\theta$ scan of the (020) Bragg peak for Ta$_2$Ni(Se$_{1-x}$S$_x$)$_5$ family of materials at room temperature.
	}
	\label{fig:XRD}
\end{figure}

In Fig.~\ref{fig:XRD} we show the $2\theta$ scan for the (020) Bragg peak measured at the room temperature with 0.028$^\circ$ spectral resolution. The full widths at half maximum (FWHM) of the peaks for \TNS and \TNSS is comparable, about $0.06^\circ$, similar or smaller than the values for \TNS single crystals measured by other groups \cite{Transport2017,nakano2018} (around $0.1^\circ$ for most peaks in samples  used for structural refinement \cite{nakano2018}). For both alloy compositions the Bragg peak is only 10\% wider, without signs of splitting, that indicates weak influence of the chemical randomness on the crystalline structure and excludes the presence of gross inhomogeneities or phase separation in alloy crystals. 
The position of the peaks evolves continuously with sulfur content $x$. 

The characterization of the crystals by temperature dependence of resistance and by transmission electron microscopy (TEM) is given in the accompanying article~\cite{Ye2021}. 
These analyses provide complementary evidences for the phase transition temperature. 

The quality of crystals is further demonstrated by the small residual linewidth for all phonon modes. 
We fitted the temperature dependence of the half width at half maximum (HWHM) of these modes by a standard model assuming anharmonic decay into two phonons with identical frequencies and opposite momenta~\cite{Klemens1966}:
\begin{equation}
\gamma_p(T)=\gamma_0+\gamma_2[1+\frac{2}{e^{\hbar\omega_0/2k_B T}-1}],
\label{eq:gammaTwo}
\end{equation}
where $\gamma_2$ is due to phonon nonlinear effects, while $\gamma_0$ characterizes the residual temperature independent scattering due to structural imperfections.
We find that the value of $\gamma_0$ for the lowest-energy phonon constitutes only 1.4\%, 2.8\%, 2.2\% and 1.4\% for $x=0,0.25,0.67,1$ compositions, respectively~\cite{Ye2021}. The small ratios indicates that the structural imperfections do not affect the lattice dynamics even at the lowest measured frequencies.  
Full account of the phonon behaviors in \TNSX is presented in the accompanying article~\cite{Ye2021}. 
In particular, for the A$_{g}$-symmetry phonon modes, the $\gamma_0$ values for the crystals used in this study~\cite{Ye2021} are less by 30-50\% than the values reported in another Raman study~\cite{Raman2019}.

\subsection{Raman measurements}

The samples used for Raman study were cleaved in ambient conditions to expose the $ac$ plane; the cleaved surfaces were then examined under a Nomarski microscope to find a strain-free area. Raman-scattering measurements were performed in a quasi-back-scattering geometry from the samples mounted in a continuous helium gas flow cryostat.

We used a custom fast f/4 high resolution 500/500/660\,mm focal lengths triple-grating spectrometer for data acquisition. For acquisition of the low-frequency Raman response, we used 1800\,mm$^{-1}$ master holographic gratings; three slit configurations were used: 100\,$\mu$m slit width providing 0.19\,meV spectral resolution; 50\,$\mu$m slit width rendering 0.10\,meV spectral resolution; 25\,$\mu$m slit width rendering 0.06\,meV spectral resolution. For the high-frequency Raman response, we used 150\,mm$^{-1}$ ruled gratings and 100\,$\mu$m slit width, providing 2.8\,meV spectral resolution. All the data were corrected for the spectral response of the spectrometer. 
The measured intensity was normalized by the incident laser power and the collection time. 

For polarization optics, a Glan-Taylor polarizing prism (Melles Griot) with a better than 10$^{-5}$ extinction ratio to clean the laser excitation beam and a broad-band 50\,mm polarizing cube (Karl Lambrecht Corporation) with an extinction ratio better than 1:500 for the analyzer was used. 
This implies that the instrumental "leakage" not related to the actual breaking of the crystalline symmetries is smaller than 0.2\% in our setup.

The 647 and 676\,nm line from a Kr$^+$ ion laser was used for excitation. 
Incident light was focused to an elongated along the slit direction 50$\times$100\,$\mu$m$^{2}$ spot. 
For data taken below 310\,K, laser power of 8\,mW was used. To reach temperature above 310\,K, we kept the environmental temperature at 295\,K and increased laser power to reach higher sample temperature in the excitation spot. All reported data were corrected for laser heating in two mutually consistent ways \cite{Ye2021}: (i) by Stokes/anti-Stokes intensity ratio analysis, based on the principle of detailed balance, (ii) by checking laser power that is inducing the phase transition. In addition to that, we performed a thermoconductivity model calculation that suggests a linear scaling of the heating rate with the beam spot size.

The measured secondary-emission intensity $I(\omega,T)$ is related to the Raman response $\chi''(\omega,T)$ by $I(\omega,T)=[1+n(\omega,T)]\chi''(\omega,T)$, where $n$ is the Bose factor, $\omega$ is energy, and $T$ is temperature. 

\subsection{The error analyses}

As for the error analysis pertaining to Fig.\,2(a) of main text, we performed each measurement four times and used the standard deviation as an estimation of the shot noise of each data point. 
After subtracting a linear background we fitted the phonon modes with a Lorenzian lineshape, with each data point weighted by its standard deviation, to find the uncertainty of the mode intensity.
We then calculated the uncertainty of the ratio shown in Fig.\,2(a) of main text by error propagation formula. 

The error bars shown in Fig.\,2(b) of main text represents one standard deviation from the Fano fits [Section.~\ref{Fano}] of the spectra with each data point carrying its own weight. 

The error bars of T$_{ex}$ and T$_{comb}$ shown in Fig.\,3 of main text represents one standard deviation from the linear fits in Fig.\,2(b) of main text. 
The error bars of T$_{c}$ shown in Fig.\,3 were estimated from the appearance of sharp spectral features in the "forbidden" scattering geometry; see SubSection.~\ref{S} for examples related to \TNSS, and find full account in the accompanying article~\cite{Ye2021}.

\subsection{The scattering polarization analyses}

The relationship between the scattering geometries and the symmetry channels~\cite{Hayes2004} is given in Table~\ref{tab:Exp1}. 
The polarization selection rules in Raman scattering (i.e. excitations of which symmetry are allowed to appear for given polarization geometry) are determined by the point-group symmetry of the material. 
While the intensities of the modes are not determined by the selection rules, their appearance in “forbidden” polarization geometries is an unambiguous evidence that the symmetry of the system is reduced.  
In particular, for \TNSX in the high-temperature orthorhombic phase (point group D$_{2h}$) the A$_g$ and B$_{2g}$ modes are distinct: the B$_{2g}$ modes are forbidden in $aa$ scattering geometry and the A$_g$ modes are forbidden in $ac$ scattering geometry. 

The appearance of modes, ascribed to the B$_{2g}$ symmetry in the high temperature D$_{2h}$ phase, in $aa$ scattering geometry (and those of A$_g$ symmetry – in $ac$ geometry) is a direct indication of broken mirror symmetries in the low temperature phase, leading to lowering of the point group symmetry to C$_{2h}$ such that A$_g$ and B$_{2g}$ [D$_{2h}$] representations merge in the C$_{2h}$ group. 
We emphasize that the analysis uniquely identifies the two broken mirror-plane symmetry operations, a$\rightarrow$-a and c$\rightarrow$-c, and the point group symmetry in the low temperature phase~\cite{kung2015}.  

\begin{table}[h]
	\caption{\label{tab:Exp1}The Raman selection rules in the high-temperature orthorhombic (point group D$_{2h}$) and low-temperature monoclinic (point group C$_{2h}$) phases. Upon the reduction of symmetry from D$_{2h}$ to C$_{2h}$, the A$_{g}$ and B$_{2g}$ irreducible representations of D$_{2h}$ group merge into the A$_{g}$ irreducible representation of C$_{2h}$ group.}
	\begin{ruledtabular}
		\begin{tabular}{ccc}
			Scattering&Symmetry Channel&Symmetry Channel\\
			Geometry&(D$_{2h}$ group)&(C$_{2h}$ group)\\
			\hline
			$aa$&A$_{g}$&A$_{g}$\\
			$ac$&B$_{2g}$&A$_{g}$\\
		\end{tabular}
	\end{ruledtabular}
\end{table}

\section{Fano model fitting\label{Fano}}

\subsection{Details of the Fano model}

To deduce the parameters of the excitonic and phonon modes above $T_c$, we have analyzed the  $ac$ Raman susceptibility using an extended Fano model \cite{volkov2020}. Here we provide the necessary details and demonstrate the excellent quality of the resulting fits; more detailed account can be found in Refs.\,\onlinecite{volkov2020,Ye2021}. 

The strongly asymmetric features, see Fig.\,1(m)-(o) of the main text, are indeed familiar from spectroscopic literature \cite{fano1961,klein1983} and indicate an interaction between sharply resonant excitations, three $B_{2g}$ phonons in our case, and a broad continuum of excitations. The latter is expected to arise from the overdamped excitonic fluctuations, with the dynamics governed by Eq.\,(1), first line, of the main text. 
This form follows from the time-dependent Ginzburg-Landau \cite{goldenfeld2018} description of the dynamics of the electronic order parameter $\varphi$, which can be applied to a diverse range of systems from charge-density wave \cite{yusupov2010,zong2019} to superconducting \cite{Schuller2006} ones. In particular, the resulting overdamped dynamics is consistent with the Landau damping of excitons in a semimetal.

\begin{figure}
	\centering
	\includegraphics[width=0.3\linewidth]{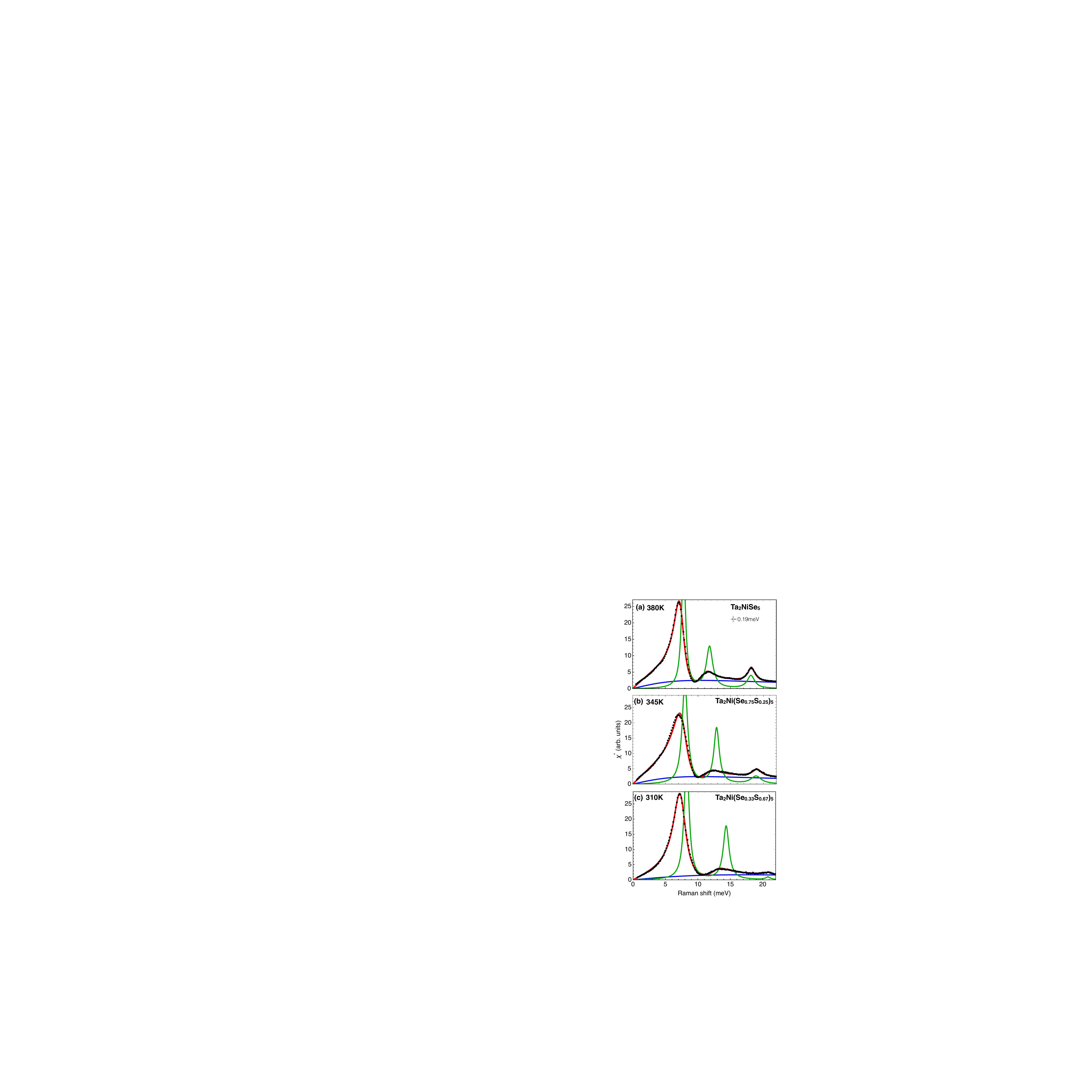}
	\caption{The low-energy Raman response (black dots) measured in ac scattering geometry above $T_c$ for samples with $x=0,0.25,0.67$. Red lines show the fits to the data with a generalized Fano model, Eqs. (\ref{eq:fanoG}-\ref{eq:fanoV}), of three phonons coupled to a continuum with a relaxational response. The deduced bare phonon and continuum responses are shown by green line and blue line with shading, respectively. Full Fano model response (red) does not equal sum of the two due to the presence of interaction $v_i$. Spectral resolution is shown in top right corner of the upper panel. 
		Data is adapted from the accompanying article~\cite{Ye2021}.}
	\label{fig:Fanofit}
\end{figure}

The general result following from Fermi's Golden Rule \cite{klein1983} is:
\begin{equation}
\chi^{\prime\prime}(\omega) = {\rm Im} \sum_{ij} T_iG_{ij}(\omega)T_j~,
\label{eq:fanoChi}
\end{equation}
where $T=\left(\begin{array}{cccc}t_{p1}&t_{p2}&t_{p3}&t_{e}\end{array}\right)$ are the matrix elements of the Raman light scattering process (for the three $B_{2g}$ phonons $t_{p1,2,3}$ and excitonic continuum, $t_{e}$, respectively) and $G$ is defined by
\begin{equation}
G=(G_0^{-1}-V)^{-1}.
\label{eq:fanoG}
\end{equation}
where
\begin{equation}
G_0=\begin{pmatrix}G_{p1}&0&0&0\\0&G_{p2}&0&0\\0&0&G_{p3}&0\\0&0&0&G_{e}\end{pmatrix}~,
\label{eq:fanoG0}
\end{equation}
\begin{equation}
V=\begin{pmatrix}0&0&0&v_1\\0&0&0&v_2\\0&0&0&v_3\\v_1&v_2&v_3&0\end{pmatrix}~.
\label{eq:fanoV}
\end{equation}
Here, $G_{\mathrm{e/p}1,2,3}(\omega)$ are the Green's functions of the electronic continuum and of the phonon mode. $v_{1,2,3}$ is the interaction matrix element between them which is approximated by a frequency-independent constant (relation between $v_i$ and $\tilde{v}_i$ in Eq. (1) of the main text will be clarified below). The form of $G$ corresponds to classical linear response functions that can be obtained by solving Eq. (1) of the main text.

Consequently, for the bare phonon Green's function, we use
\begin{equation}
G_{\mathrm{p}i}(\omega)=-\left(\frac{1}{\omega-\omega_{\mathrm{p}i}+i\gamma_{\mathrm{p}i}}-\frac{1}{\omega+\omega_{\mathrm{p}i}+i\gamma_{\mathrm{p}i}}\right)~,
\label{eq:chiphon}
\end{equation}
which corresponds, up to a proportionality constant $1/(2\omega_{\mathrm{p}i})$, to Eq. (1) of the main text in the relevant limit $\gamma_{\mathrm{p}i}\ll\omega_{\mathrm{p}i}$. The proportionality constant also results in the relation 
\begin{equation}
v_i= \frac{\tilde{v}_i}{\sqrt{2\omega_{pi}(T)}}
\label{eq:vdef}
\end{equation}
for the exciton-phonon coupling in Eq. (1) of the main text. The parameters $\omega_{\mathrm{p}i}$ and $\gamma_{\mathrm{p}i}$ respectively have the physical meaning of the mode energy and the half width at half maximum (HWHM). Finally, the excitonic continuum response takes the form:
\begin{equation}
G_{\mathrm{e}}^{-1}(\omega) = -i\omega+\Omega_{\mathrm{e}}.
\label{eq:chiel}
\end{equation}	

In the absence of interaction $v_{1,2,3}=0$, the Raman susceptibility reduces to the sum of the individual components given by $t_{\mathrm{e/p}1,2,3}^2G_{\mathrm{e/p}1,2,3}(\omega)$, with sharp peak from phonon modes superimposed on the excitonic background. However, presence of interaction leads to a dramatic asymmetric broadening of the shapes, allowing to capture the features of the experimental data.

In Fig. \ref{fig:Fanofit} (red lines) we present the fits obtained using Eqs. (\ref{eq:fanoG}-\ref{eq:fanoV}), as well as the bare excitonic (blue line) and phononic (green line) responses as:
\begin{equation}
\chi_{\mathrm{\mathrm{opt}}}''(\omega) = \sum_{i=1}^3 t_{\mathrm{p}i}^2 {\rm Im} G_{\mathrm{p}i}(\omega);
\;
\chi_{\mathrm{cont}}''(\omega)  = \frac{t_{\mathrm{e}}^2 \omega}{\omega^2 + \Omega_{\mathrm{e}}^2}.
\label{eq:chibare}
\end{equation}
One observes that the fit quality away from $T_{\mathrm{c}}$ is excellent for all S contents. The resulting value of $\Omega_e(T)$ are shown in Fig. 2 (b) of the main text. The detailed values of the other parameters are reported in the accompanying Article \cite{Ye2021}.

\subsection{Phase diagram}
We discuss now how the phase diagram in Fig. 3 of the main text was obtained. The purely excitonic transition temperature is defined by the excitonic mode having zero energy $\Omega_e(T_{ex}) = 0$ (Fig. 2(b) of the main text). One observes, however, that it precisely coincides with the condition for the static ($\omega=0$) electronic susceptibility to diverge:
\begin{equation}
\chi_{cont}(\omega=0,T) = \frac{t_e^2}{\Omega_e(T)};\;\chi_{cont}(\omega=0,T_{ex})\to\infty,
\end{equation}
corresponding to a thermodynamic phase transition in a purely excitonic system (i.e. ignoring phonons).
\\
In the presence of optical phonons, the static response of the system is given by 
\begin{equation}
\chi(\omega=0,T) = \sum_{ij} T_iG_{ij}(\omega=0,T)T_j.
\end{equation}
the thermodynamic transition $\chi(\omega=0,T)\to\infty$ occurs at $T_{comb}$. It can be shown that (see also \cite{volkov2020,Ye2021}) this occurs when the determinant of $G_{ij}^{-1}(\omega=0,T)$ vanishes, i.e.
\begin{equation}
\Omega_e(T_{comb}) -\sum_i \frac{2 v_i^2}{\omega_{pi}(T_{comb})} \equiv \Omega^{comb}_e(T_{comb}) = 0.
\end{equation}
This corresponds with the condition derived from the Eq. (1) of the main text after substituting (see Eq. (\ref{eq:vdef}) above) $v_i= \tilde{v}_i/\sqrt{2\omega_{pi}(T)}$.
\\
The influence of the acoustic modes can be taken into account within the same formalism as above, by introducing an additional phonon mode into Eq. (\ref{eq:fanoV}) with energy $\omega_s=c_s q$, linewidth $\gamma_s=r_s q$  and coupling to excitons $v_s=\beta_q \sqrt{q}$, where $q$ is the wavevector that is infinitesimally small for Raman scattering (see \cite{volkov2020,Ye2021} for details). Nonetheless, even in the $q\to0$ limit, the susceptibility of the system including interaction of excitons with both acoustic and optical modes diverges only when:
\begin{equation}
\Omega^{comb}_e(T_{comb}^{str})
- \frac{2 \beta_s^2}{c_s} = 0,
\end{equation}
where $\frac{2 \beta_s^2}{c_s} \equiv \lambda^2/(2C_{ac})$. This can be generalized to the case of the presence of a ferroeleastic instability by using a temperature-dependent $C_{ac}(T)$:
\begin{equation}
C_{ac}^{-1}(T) = C_{ac(0)}^{-1}+\frac{a}{T-T_{FE}},
\end{equation}
in which $T_{FE}=120$\,K corresponds to the phase transition temperature of Ta$_2$NiS$_5$. Because $T_{comb}^{FE}$ satisfies
\begin{equation}
\Omega^{comb}_e(T_{comb}^{FE}) - \frac{\lambda^2}{2C_{ac}(T_{comb}^{FE})} = \Omega^{comb}_e(T_{comb}^{FE}) - \frac{\lambda^2}{2C_{ac(0)}}-\frac{\lambda^2a}{2(T_{comb}^{FE}-T_{FE})} = 0,
\label{eq:MaiRev}
\end{equation}
we obtain the values of $\lambda^2C_{ac(0)}^{-1}$ and $\lambda^2a$ by fitting the phase transition temperatures of Ta$_2$Ni(Se$_{1-x}$S$_x$)$_5$ family with Eq.\,\eqref{eq:MaiRev}, and then use Eq.\,\eqref{eq:MaiRev} to obtain $T_{comb}^{FE}$ (Fig. 4, black line).

\section{Spectral results for {{Ta$_2$NiS$_5$}}}

\subsection{The exciton mode}

\begin{figure}[h]
	\centering
	\includegraphics[width=0.4\linewidth]{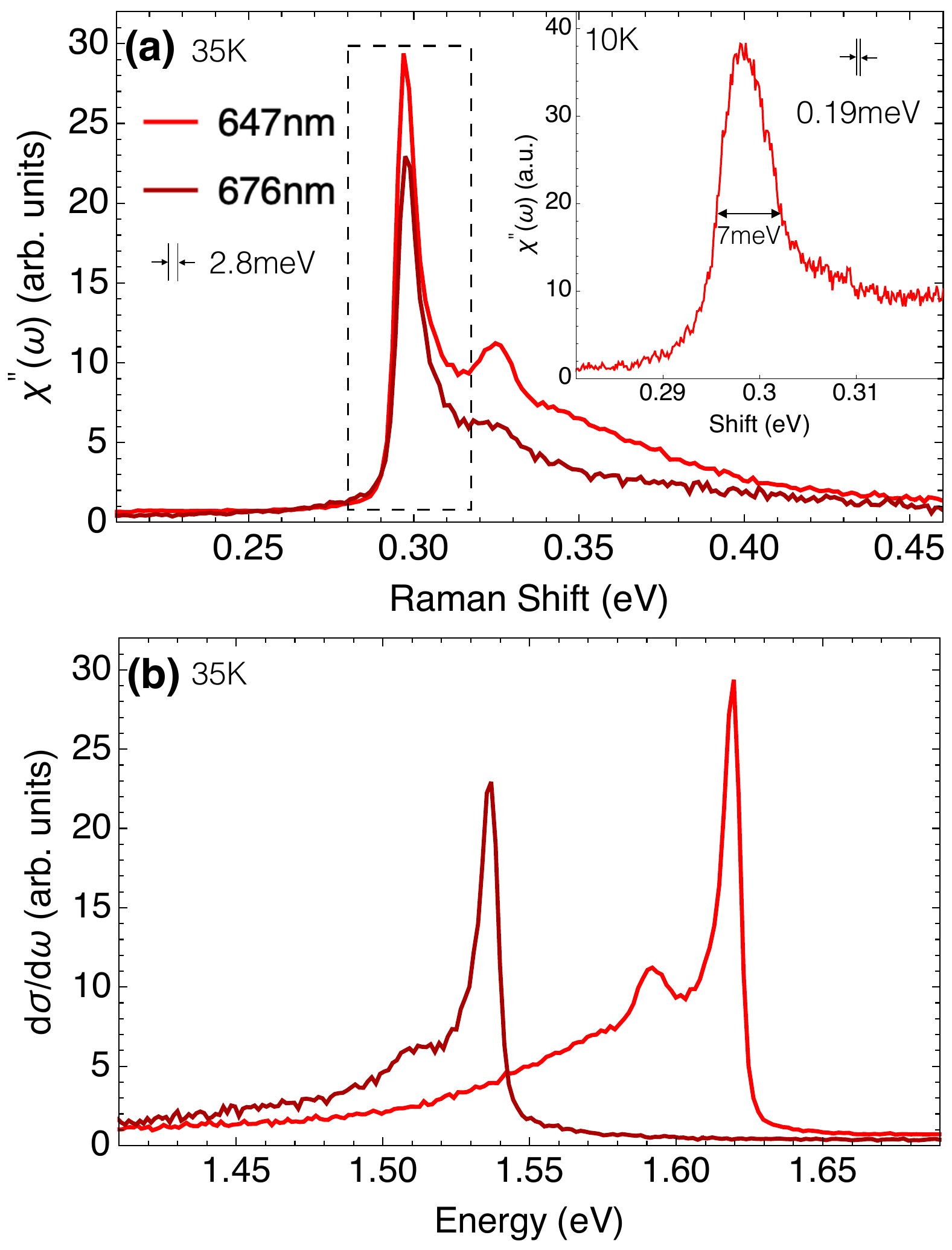}
	\caption{The high-energy spectra of Ta$_2$NiS$_5$ measured in the $ac$ scattering geometry at 35\,K. (a) Raman response as a function of Raman shift, the energy difference between the laser-photon energy and the scattered-photon energy. The inset of (a) shows the high-resolution spectrum of the exciton mode at 10\,K, corresponding to the energy range enclosed by the dashed box in (a). The full width at half maximum (FWHM) of the exciton mode is 7\,meV. (b) Differential cross section as a function of scattering-photon energy.}
	\label{fig:Exciton}
\end{figure}

In Fig.\,\ref{fig:Exciton}(a) we compare the exciton mode of Ta$_2$NiS$_5$ measured by two different excitation wavelengths at 647 and 676\,nm. 
The Raman shift of this mode at about 300\,meV is excitation-independent, indicating that the mode is a genuine Raman feature. In Fig.\,\ref{fig:Exciton}(b) we show, for comparison, the scattering cross section plotted against absolute photon energy. 
We interpret the weaker spectral feature at 325\,meV as the second state of the Rydberg series, the 2S exciton, which exhibits less intensity than the 1S exciton at 300\,meV. 
The apparent continuum of excitations up to about 400\,meV can then be attributed, in analogy with optical absorption spectroscopy, to the interband transitions \cite{elliot1957,haug2009quantum} with possible contributions from phonon-assisted exciton transitions \cite{toyozawa1964,segall1968} to finite-momentum exciton states as well as Rydberg states of higher order \cite{elliot1957,haug2009quantum}. 

\subsection{The structural phase transition\label{S}}

\begin{figure}[t]
	\centering
	\includegraphics[width=0.6\linewidth]{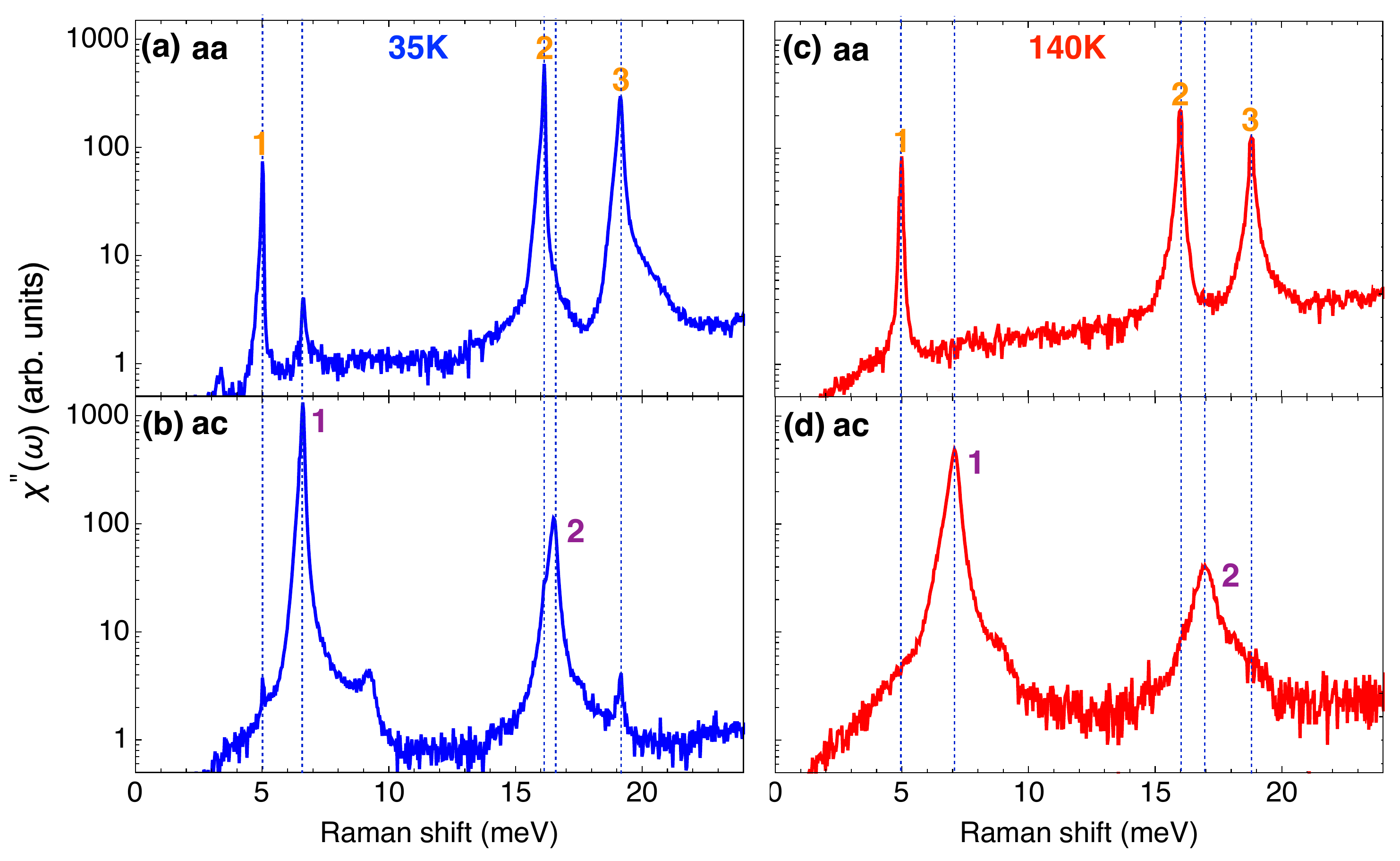}
	\caption{The low-energy phonon spectra of Ta$_2$NiS$_5$. (a) Raman response in the aa geometry at 35\,K. (b) Raman response in the ac geometry at 35\,K. (c) Raman response in the aa geometry at 140\,K. (d) Raman response in the ac geometry at 140\,K. The A$_{g}$ phonon modes are labelled with orange numbers and the B$_{2g}$ modes with purple numbers.}
	\label{fig:LeakageP}
\end{figure}

As shown in Table~\ref{tab:Exp1}, due to the breaking of two mirror-plane symmetry operations below T$_c$, the A$_{g}$ and B$_{2g}$ representations of the high-temperature phase, point group D$_{2h}$, merge into A$_{g}$ representations of the point group C$_{2h}$. 
Thus, the orthorhombic-to-monoclinic structural phase transition removes the orthogonality between A$_{g}$ and B$_{2g}$ [D$_{2h}$] representations, making the phonon modes observable in "forbidden" scattering geometry, namely, the A$_{g}$-symmetry modes start to appear in the $ac$ geometry, and the B$_{2g}$-symmetry modes begin to appear in the $aa$ geometry below the transition temperature. 
In Fig.~\ref{fig:LeakageP} (a-b) we present the phonon spectra of Ta$_2$NiS$_5$ at 35\,K in the two scattering geometries. 
The A$_{g}^{(1)}$, A$_{g}^{(3)}$, and B$_{2g}^{(1)}$ modes appears in the "forbidden" scattering geometry as a distinct spectral peak, while the A$_{g}^{(2)}$ and B$_{2g}^{(2)}$ modes show a weak shoulder feature in the "forbidden" scattering geometry. 
On the contrary, there are no such spectral features in the orthogonal scattering geometry at 140\,K [Fig.\,\ref{fig:LeakageP}(c-d)]. 
Such difference clearly indicates the broken symmetry due to structural phase transition for Ta$_2$NiS$_5$. Moreover, the fact that $aa$ can $ac$ scattering geometries become no longer orthogonal below the phase transition temperature indicates that the two broken mirror-plane symmetry operations specifically are a$\rightarrow$-a and c$\rightarrow$-c. 

\begin{figure}[h]
	\centering
	\includegraphics[width=0.27\linewidth]{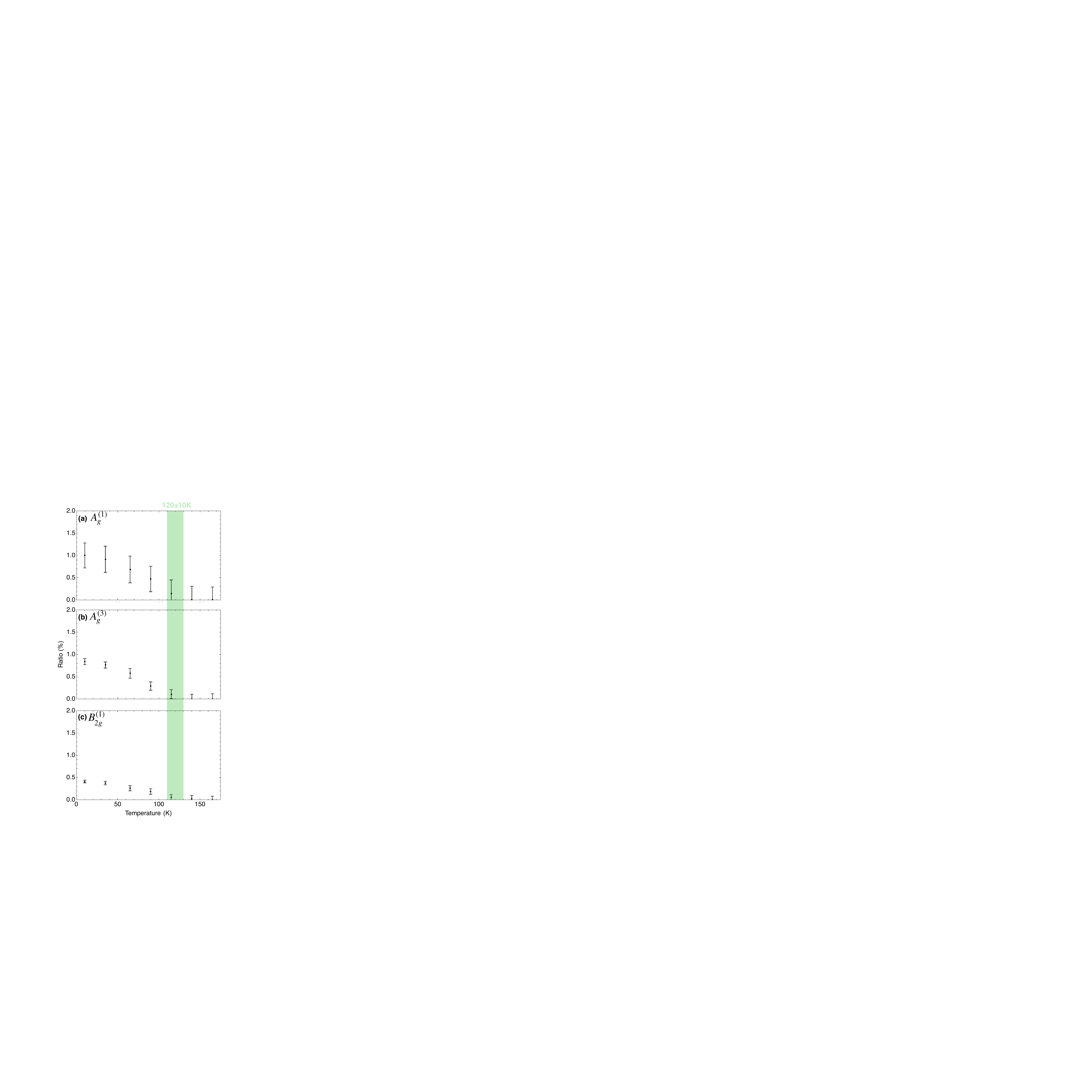}
	\caption{Temperature dependence of the intensity in the "forbidden" scattering geometry for the three phonon modes (labelled in Fig.~\ref{fig:LeakageP}), normalized by their respective intensity in the dominant scattering geometry.}
	\label{fig:IntensityP}
\end{figure}

Within the Landau theory, the intensity of the modes in the "forbidden" scattering geometry is proportional to the square of the order parameter, which can be represented by, e.g., the deviation of the angle $\beta$ between $a$ and $c$ axes from 90$^\circ$. 
It is therefore expected that such intensity remains zero above the transition temperature, and starts to increase monotonically on cooling below the transition temperature. 
The effect of fluctuations might lead to a finite intensity in the "forbidden" scattering geometry in a narrow temperature range above the transition temperature. 
In Fig.\,\ref{fig:IntensityP} we show the temperature dependence of the intensity in the "forbidden" scattering geometry for the A$_{g}^{(1)}$, A$_{g}^{(3)}$, and B$_{2g}^{(1)}$ modes~\cite{Ye2021}. 
As the mode's appearance in the "forbidden" scattering geometry implies the onset of symmetry breaking, we determine the transition temperature for Ta$_2$NiS$_5$ to be 120$\pm$10\,K.

\begin{figure}[h]
	\centering
	\includegraphics[width=0.6\linewidth]{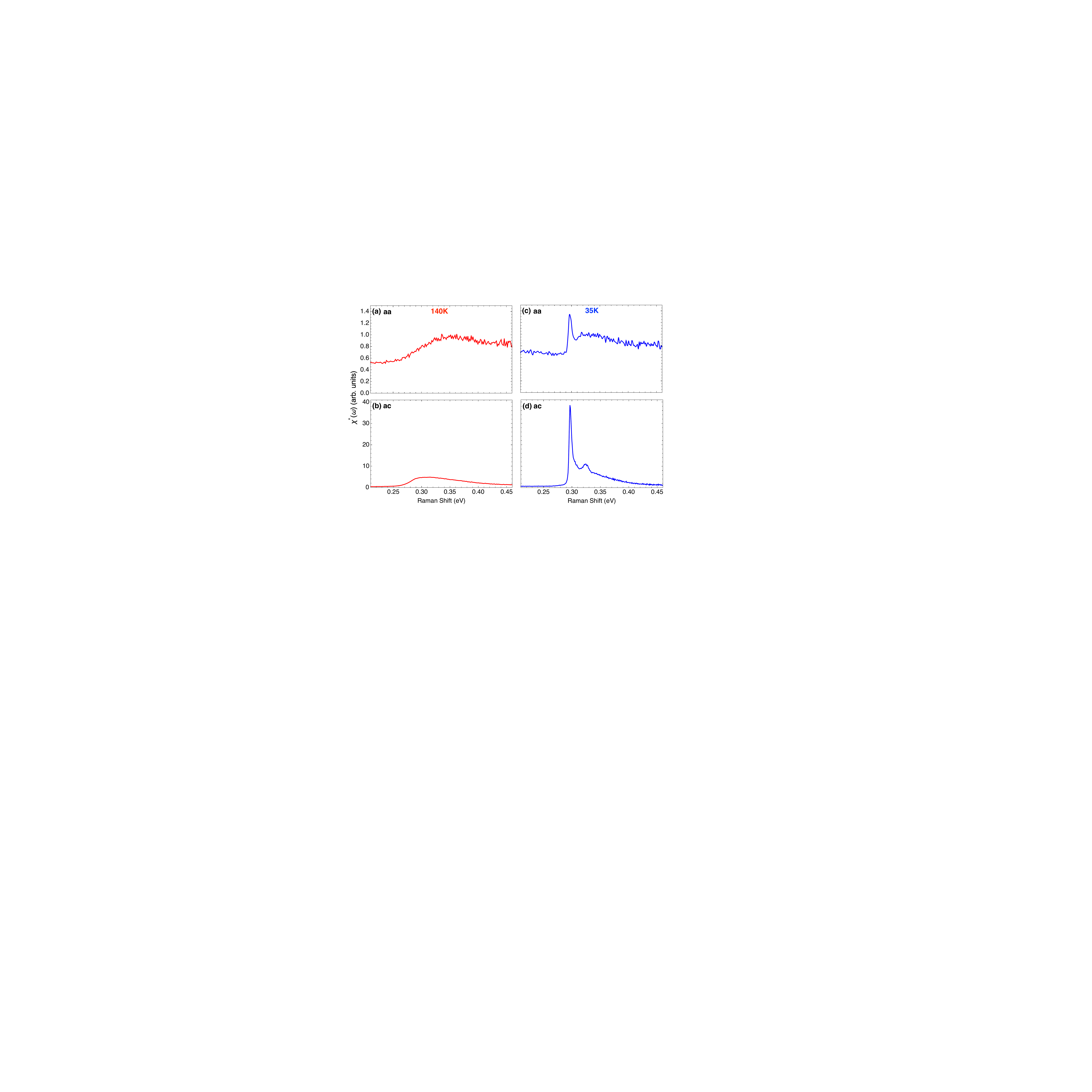}
	\caption{The high-energy spectra of Ta$_2$NiS$_5$. (a) Raman response in the aa geometry at 35\,K. (b) Raman response in the ac geometry at 35\,K. (c) Raman response in the aa geometry at 140\,K. (d) Raman response in the ac geometry at 140\,K.}
	\label{fig:LeakageE}
\end{figure}

\begin{figure}[h]
	\centering
	\includegraphics[width=0.4\linewidth]{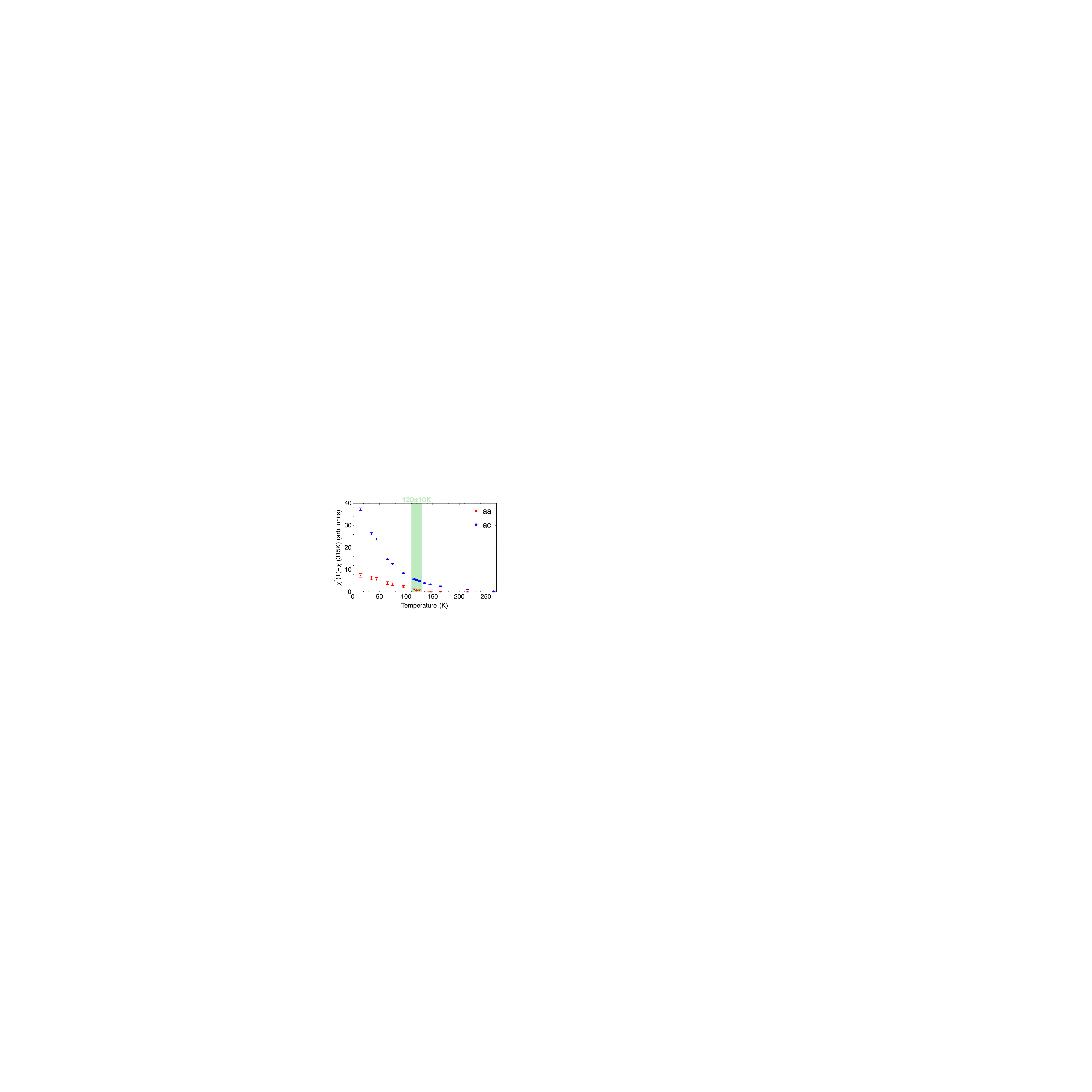}
	\caption{Temperature dependence of the Raman response $\chi^{\prime\prime}$(T) at the peak energy of the exciton mode (0.298\,eV) for Ta$_2$NiS$_5$, with the value measured at 315\,K, $\chi^{\prime\prime}$(315\,K), subtracted.}
	\label{fig:IntensityE}
\end{figure}

The symmetry breaking not only allows the phonon modes to appear in the "forbidden" scattering geometry, but also has an influence on the electronic excitations. Hence, the high-energy B$_{2g}$-symmetry [in the high-temperature D$_{2h}$ point group] exciton mode should emerge in the $aa$ scattering geometry at low temperature as well. 
This is indeed the case, as shown in Fig.\,\ref{fig:LeakageE}. 
To quantify this effect as a function of temperature, we show in Fig.\,\ref{fig:IntensityE} the Raman response at the exciton peak energy as a function of temperature. 
For both geometries we have subtracted the value at 315 K. 
One observes that in the $ac$ geometry the intensity grows on cooling at all temperatures, although the increase is faster below 120\,K. 
In the $aa$ geometry, on the other hand, the intensity does not evolve at all between 315\,K and 120\,K, consistent with the energy being below that of the interband transition. 
Below 120\,K, the intensity starts to grow in the "forbidden" for this B$_{2g}$-symmetry exciton $aa$ scattering geometry, consistent with the symmetry-breaking transition temperature deduced from the "leakage" of phonon modes. 

\begin{figure}[h]
	\centering
	\includegraphics[width=0.4\linewidth]{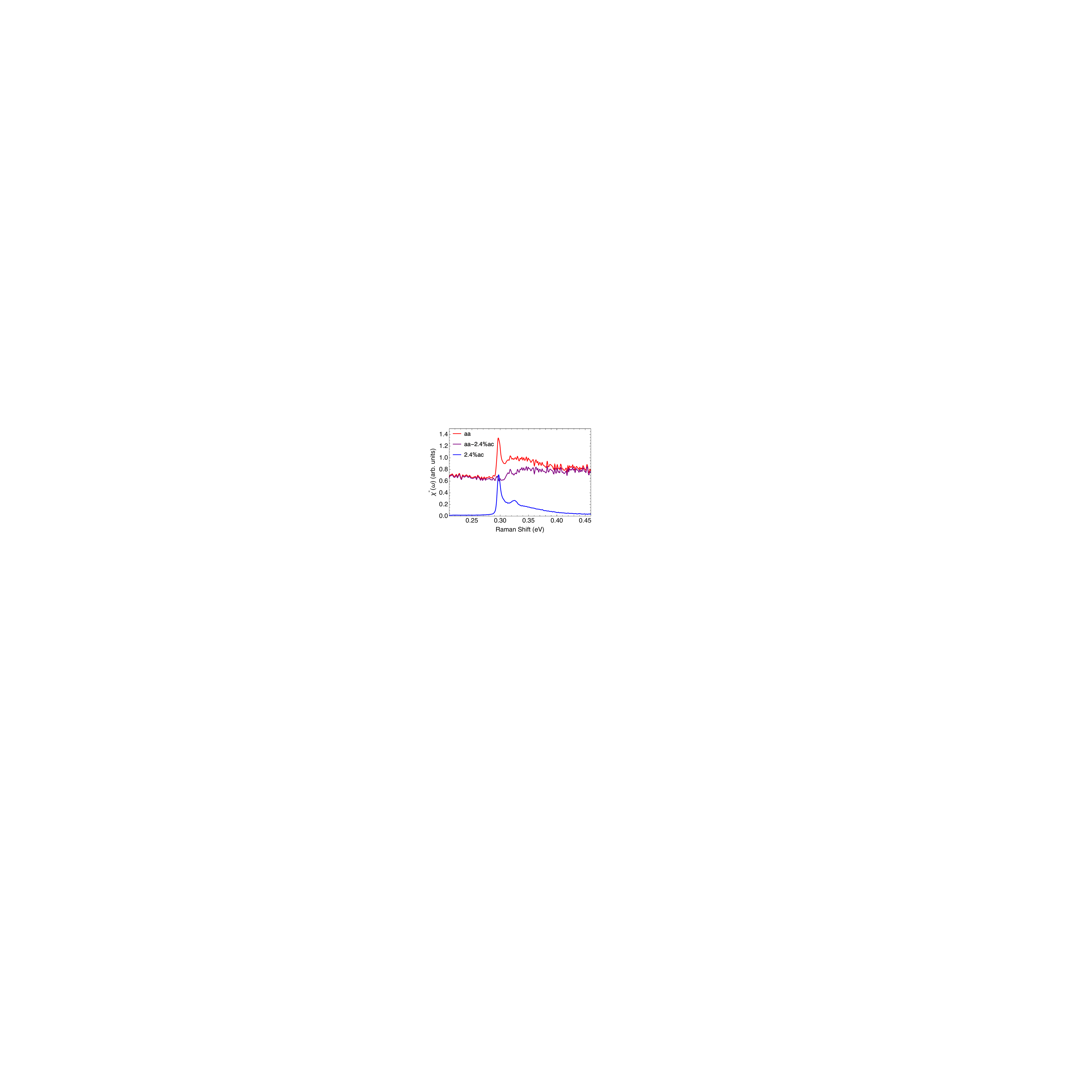}
	\caption{Analysis of the exciton-mode "leakage" at $35$\,K. Red line is the measured spectrum in the $aa$ scattering geometry; subtracting 2.4\% of the $ac$ spectrum (blue line) from the $aa$ spectrum renders an essentially featureless spectrum (purple line), indicating that there is a 2.4\% leakage of the exciton mode.}
	\label{fig:Subtract}
\end{figure}

We can also study the leakage of the whole exciton-associated feature below $T_c$. As the exciton intensity in the $ac$ scattering geometry is much larger than that in $aa$ geometry, we can neglect the leakage of $aa$ into $ac$, and consider only the former effect. To demonstrate that the weak spectral peak in the $aa$ scattering geometry results from the symmetry breaking, in Fig.\,\ref{fig:Subtract} we analyze $\chi''_{\mathrm{aa}}(\omega)- p\cdot\chi''_{\mathrm{ac}}(\omega)$, where $0<p<1$. We find that for $p=2.4\%$, the spectrum after subtraction becomes essentially featureless. This suggests all the low-temperature evolution of intensity around 0.3 eV in aa geometry can be attributed to "leakage" from ac geometry. 

\end{document}